\documentclass[aps,prd,preprint,superscriptaddress,tightenlines,nofootinbib]{revtex4}

\usepackage{graphicx}
\usepackage{dcolumn}
\usepackage{bm}

\begin{document}

\preprint{CLNS 08/2035}     
\preprint{CLEO 08-18}       

\title{Search for Lepton Flavor Violation in Upsilon Decays}

\author{W.~Love}
\author{V.~Savinov}
\affiliation{University of Pittsburgh, Pittsburgh, Pennsylvania 15260}
\author{A.~Lopez}
\author{S.~Mehrabyan}
\author{H.~Mendez}
\author{J.~Ramirez}
\affiliation{University of Puerto Rico, Mayaguez, Puerto Rico 00681}
\author{G.~S.~Huang}
\author{D.~H.~Miller}
\author{V.~Pavlunin}
\author{B.~Sanghi}
\author{I.~P.~J.~Shipsey}
\author{B.~Xin}
\affiliation{Purdue University, West Lafayette, Indiana 47907}
% ...
\collaboration{CLEO Collaboration}
\noaffiliation

\date{July 17, 2008}

\begin{abstract} 
In this Letter we describe a search for 
lepton flavor violation (LFV) in the bottomonium system. 
We search for leptonic decays 
$\Upsilon(nS) \to \mu \tau$ ($n=1,2$~and $3$) 
using the data collected with the CLEO~III detector. 
We identify the $\tau$ lepton using its leptonic decay 
$\nu_\tau \bar{\nu}_e e$ and utilize multidimensional 
likelihood fitting with PDF shapes measured from 
independent data samples. We report our estimates of 
95\% CL upper limits on LFV branching fractions of $\Upsilon$ 
mesons. We interpret our results in terms of the exclusion 
plot for the energy scale of a hypothetical new interaction versus 
its effective LFV coupling in the framework of effective field theory. 
\end{abstract}

\pacs{11.30.Fs, 12.60.-i, 13.20.Gd}
\maketitle

The subject of this Letter is a search for lepton flavor violating (LFV) 
bottomonium decays $\Upsilon(nS) \to \mu \tau$ ($n=1,2$~and~$3$). 
Such decays are predicted by various theoretical models that 
allow tree-level flavor-changing neutral currents (FCNC), 
including, {\it e.g.}, $R$-parity violating and 
large $\tan{\beta}$ ~SUSY scenarios, 
leptoquarks, and other models inspired by the idea of 
grand unification \cite{pati,georgi}. 
Our search is motivated by the discovery of large mixing between 
the second and the third generations in the neutrino sector \cite{sno}. 

The conservation of lepton, lepton flavor, and baryon quantum 
numbers in the standard model (SM) is due to accidental global 
symmetries of its Lagrangian. All such symmetries should be 
violated at higher energies, where we expect the emergence of 
a gauge group of the higher-order symmetry 
that presumably describes fundamental interactions at 
the energy scale of grand unification. 
The search for beyond the standard model (BSM) physics 
in low-energy processes is facilitated by parameterizing 
such BSM physics, 
without explicitly invoking its unknown dynamics, 
in the framework of the Wilson operator product expansion (OPE) 
and effective field theory. 
The large lepton mass hierarchy and dimensional analysis 
suggest that the effects of BSM physics are most likely to be observed in 
transitions that involve heavy quarks, muons, and $\tau$ leptons. 
In the OPE the effects of BSM physics 
in decays $\Upsilon(nS) \to \mu \tau$ 
are expressed by the four-fermion diagonal operators \cite{silagadze,sher} 
that respect the full electroweak SM gauge group 
$SU(2)_L \bigotimes U(1)_Y$
and contribute to the SM Lagrangian as 

\begin{equation}
{\cal L}_{\mathrm{eff}} = {\cal L}_{\mathrm{SM}} + \frac{4 \pi \alpha_N}{\Lambda^2} (\bar{\mu} \Gamma_\mu \tau)(\bar{b} \gamma^\mu b),
\label{eq_1}
\end{equation}

\noindent where $\Gamma_\mu$ is a vector ($\gamma_\mu$) or an axial ($\gamma_\mu \gamma_5$) 
current or their combination, 
$\Lambda$ is the scale of BSM physics 
and 
$\alpha_N$ is the effective LFV coupling of the new gauge symmetry associated with BSM.  

Previously, we searched for LFV in $B$ meson decays \cite{cleo_lfv_b},  
while the BES experiment searched for LFV in $J/\psi$ decays \cite{bes_lfv}. 
Those two analyses probed the BSM contributions 
parameterized by the operators 
$(\bar{\mu} \Gamma \tau)(\bar{b} \Gamma d)$ 
($\Gamma = \gamma_5, \gamma_5 \gamma_\mu$) 
and
$(\bar{\mu} \Gamma_\mu \tau)(\bar{c} \gamma^\mu c)$, 
respectively. 
In the analysis presented in this Letter we probe the four-fermion operators 
$(\bar{\mu} \Gamma_\mu \tau)(\bar{b} \gamma^\mu b)$. 

The CLEO~III detector, centered on the interaction region 
of the Cornell Electron Storage Ring (CESR), 
is a versatile multi-purpose particle detector \cite{cleo-iii}. 
Relevant components of the apparatus include 
a nearly $4\pi$ tracking volume surrounded by 
a Ring Imaging Cherenkov Detector (RICH) \cite{cleo-rich}, 
an electromagnetic CsI(Tl) crystal calorimeter, 
and a muon identification system \cite{cleo-muons} 
consisting of proportional wire chambers 
that provide two-dimensional position information. 
The tracking volume, located inside an axial magnetic field 
of 1.5~T, is instrumented with a 47-layer wire drift chamber 
and a four-layer silicon strip detector that allow us to measure 
the positions, momenta, and specific ionization energy losses ($dE/dx$) of 
charged particles with momentum resolution of 0.35\% (0.86\%) at 1~GeV/$c$ (5~GeV/$c$)
and a $dE/dx$ resolution of 6\%. 
The calorimeter, first installed in the CLEO~II detector \cite{cleo-ii}, 
forms a cylindrical barrel around the tracking volume 
and has resolution of 2.2\% (1.5\%) for 1~GeV (5~GeV) photons and electrons. 
The calorimeter, just inside the magnet coil, 
is followed by 
Fe flux-return plates 
interleaved with 
three layers of the muon identification system. 

We search for non-SM leptonic decays 
$\Upsilon(nS) \to \mu \tau$ ($n=1,2$~and $3$) 
using the data collected with the CLEO~III detector. 
We identify the $\tau$ lepton using an electron from its leptonic decay 
$\nu_\tau \bar{\nu}_e e$. 
We use 
data samples that contain 
$20.8$, 
$9.3$, 
and 
$5.9$ 
million 
$\Upsilon(1S)$, $\Upsilon(2S)$, and $\Upsilon(3S)$ 
resonant decays, respectively \cite{cleo_data_1,cleo_data_2_and_3}.  
Integrated $e^+e^-$ luminosities of these signal data samples 
are 1.1~${\rm fb^{-1}}$, 1.3~${\rm fb^{-1}}$, and 1.4~${\rm fb^{-1}}$. 
We use the $\Upsilon(4S)$ (6.4~${\rm fb^{-1}}$) and 
hadronic ``continuum'' (2.3~${\rm fb^{-1}}$ collected 60~MeV below the $\Upsilon(4S)$ energy) 
data 
to measure the shapes of 
probability density functions (PDFs) and resolution parameters 
used in maximum likelihood (ML) signal fits described later in this Letter. 
We also use the $\Upsilon(4S)$ and continuum data to verify the overall 
reconstruction and trigger efficiency and to estimate systematic errors. 

The signature of our signal is 
a muon with $p_\mu / E_\mathrm{beam} \approx 0.97$ 
and an electron from the decay of the $\tau$ lepton. 
We select events with two reconstructed tracks of opposite electric charge. 
One track is identified as a high-quality muon candidate 
by requiring that it penetrate five hadronic interaction lengths. 
The other track should satisfy electron identification criteria 
by requiring a $\pm 3\sigma$ consistency 
with the theoretically-predicted $dE/dx$ contribution 
and $0.85 \le E/p \le 1.10$, where $E$ is the energy reconstructed in the 
region of the electromagnetic calorimeter matched to the projection 
of electron's track of momentum $p$. 
Electron and muon candidates should not also be identified as the candidates of the other lepton species. 
The beam-energy normalized momenta of the muon and electron candidates, 
$x = p_\mu/E_{\mathrm{beam}}$ and $y = p_e/E_{\mathrm{beam}}$, 
are required to be within the ranges 
$0.87 \le x \le 1.02$ and $0.10 \le y \le 0.85$. 

The geometric acceptance of tracking is $\approx 86$\% for two tracks. 
Track reconstruction efficiency for the signal is 83\% in the acceptance region. 
The muon system coverage is 84\% of the solid angle and 
the efficiency of muon identification in that region 
is 92\% per muon when its charged track is reconstructed. 
Electron identification is 95\% efficient, 
due to the calorimeter's angular acceptance. 
The trigger for signal events in fiducial region of the detector is 93\% efficient. 
The efficiency for selecting events in the $x$ and $y$ regions 
(after applying all other criteria) is 95\%. 
Trigger and reconstruction efficiency for the signal is 50\%. 
Its product with the ${\cal B}(\tau \to \nu_\tau \bar{\nu}_e e) = (17.84 \pm 0.05)$\% \cite{PDG}
yields an overall efficiency of 8.9\%. 

\begin{figure}[h]
\includegraphics*[width=6.50in]{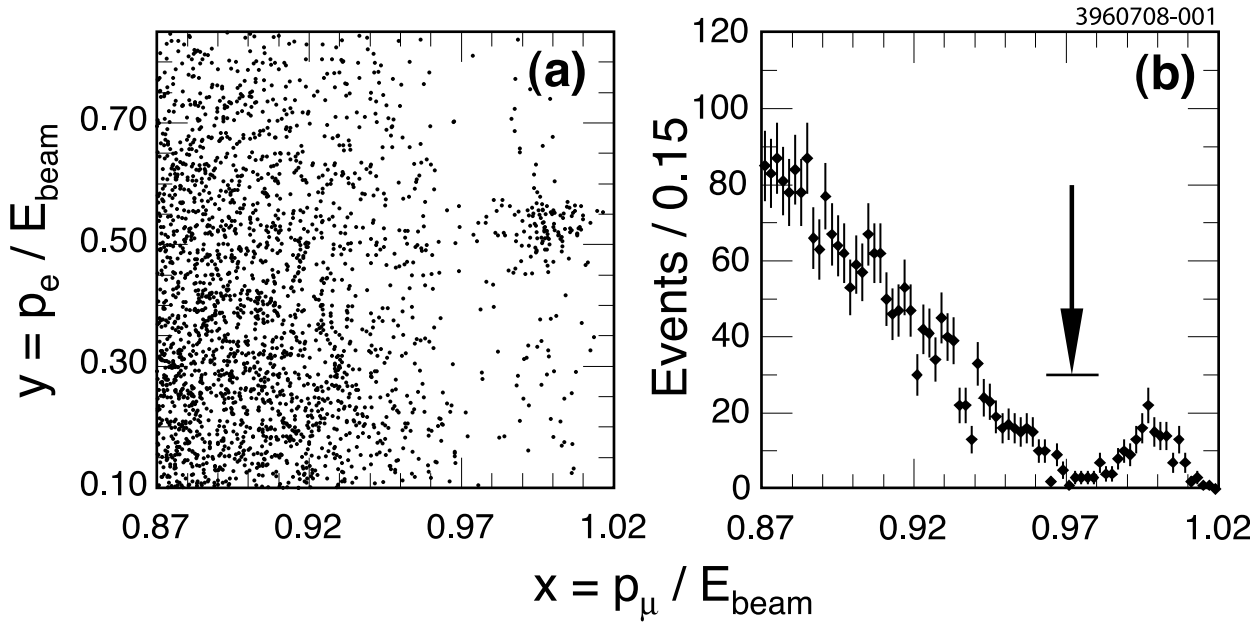}
\caption{(a) The scatter plot of $y$ versus $x$ 
and 
(b) its binned $x$ projection for calibration data. 
The location of the hypothetical signal peak is indicated by the arrow, 
where the width of the horizontal bar at its tip is $\pm \sigma(x)$. 
}
\label{fig_prl_1}
\end{figure}

We do not expect to find LFV in the $\Upsilon(4S)$ 
and hadronic continuum data used to calibrate our analysis method. 
Even if LFV BSM physics, {\it e.g.}, quantum gravity, 
becomes strong at a TeV energy scale, 
LFV would occur in dilepton decays of the $\Upsilon(4S)$ 
at a much smaller rate 
than in decays of lower-mass $b\bar{b}$ resonances, 
because 
the products of their 
production cross sections 
and SM dilepton partial widths  
are significantly larger than that for the $\Upsilon(4S)$. 
The BaBar experiment has recently published an upper limit (UL) 
for $\sigma(\tau \mu)/\sigma(e e)$ at the $\Upsilon(4S)$ energy \cite{babar}. 
Their UL suggests that less than 3 LFV events would be observed in our calibration data. 
We show the distribution of $y$ versus $x$ for our calibration data in Fig.~\ref{fig_prl_1}(a) 
and the projection onto the axis $x$ in Fig.~\ref{fig_prl_1}(b). 

According to our studies, confirmed by Monte Carlo (MC) simulation for QED processes, 
three backgrounds arising from $\mu$ and $\tau$ pairs 
contribute to the distributions shown in Fig.~\ref{fig_prl_1}. 
The $\mu$ pairs contribute in two ways,  
through radiative processes, and, also, 
when one muon decays to an electron in flight. 
The first contribution from $\mu$ pairs includes QED radiation 
at the vertex and hard bremsstrahlung in the detector. 
Such events satisfy our selection criteria 
when 
a radiative photon matches the muon track's projection to the calorimeter 
and 
muon identification fails. 
Such events cluster around $y=0.53$ because of the $E/p$ requirement, 
where $E$, for such background events, is the energy of radiative photon ($\approx E_{\mathrm{beam}}/2$) 
combined with a small amount of energy ($\approx 0.2$~GeV) deposited by the muon in the calorimeter. 

The second, and less frequent background from $\mu$ pairs appears when one muon decays in flight. 
This results in the actual electron detected in the calorimeter. 
Such events cluster near $x=1$ but scatter in $y$ between 0.10 and 0.85. 
Both background contributions from $\mu$ pairs differ from 
the hypothetical signal in $\Upsilon(nS)$ ($n=1,2$~and~$3$) data. 
The high-momentum background muon is most often produced at beam energy, $x = 1$ 
(though radiative processes introduce a long tail in the $x$ shape for this background), 
while the signal muon peaks at $x = 0.965, 0.968$,~and~$0.970$ 
for the $\Upsilon(1S)$, $\Upsilon(2S)$, and $\Upsilon(3S)$, respectively. 
Also, when a muon mimics an electron, the $E/p$ and $dE/dx$ distributions 
differ from those we expect for the real electrons. 
While the $dE/dx$ measurements do not have sufficient resolution 
to discriminate between electrons and muons on an event by event basis 
in the relevant momentum region, namely, around 2.5~GeV,  
discrimination between the signal and backgrounds on a statistical basis is possible. 

The production of $\tau$ pairs represents an irreducible background 
to our signal when both $\tau$ leptons decay leptonically, 
one to an electron and the other to a muon. 
The only variable that discriminates our signal 
from this background is $x$, 
the beam-energy normalized momentum of the signal muon candidate. 

To estimate the number of LFV decays in $\Upsilon(nS)$ ($n=1,2$~and~$3$) data 
we subject the events that pass the selection criteria to four-dimensional 
unbinned extended ML fits. For each probed data sample 
we maximize the likelihood function 

\begin{equation}
{L} = \frac{1}{N!}\exp\left({-\sum_{j}^{4} N_j}\right) \,
\prod_{i}^N \sum_{j}^{4} N_j {\cal P}_j(\{z\}_i,\{\alpha\}_j),
\label{equation_likelihood_definition}
\end{equation} 

\noindent where $N$ is the total number of data events in the fit; 
$i$ is the index for these events; 
$j$ is the index for fit contributions (the signal and the three backgrounds); 
$\{z\}_i$ is the vector of the four variables $x$, $y$, $dE/dx$ and $E/p$ for event $i$;  
$N_j$ is the fit parameter that corresponds to the numbers of events for fit contribution $j$; 
and ${\cal P}_j$ is the four-dimensional PDF 
with shape parameter vector $\{\alpha\}_j$ for fit contribution $j$. 

We utilize calibration data 
to find approximations for the PDFs, 
the values of their shape parameters 
and respective matrices of systematic errors. 
The correlations among the variables, 
especially important for the $\mu$-pair backgrounds, 
are included in the respective PDFs. 
To take into account initial state radiation, 
we parameterize the $x$ shape of the $\mu$-pair background 
using a Gaussian with a long asymmetric tail. 
The $E/p$ shape for real electrons is also parameterized by such a Gaussian.  
The $x$ shape for the $\tau$-pair background is parameterized 
by a first order polynomial smeared by Gaussian detector resolution 
measured using the data. 
The $E/p$ shape for the muon matched with a radiative photon in the calorimeter, 
therefore misidentified as the signal electron candidate, 
is approximated by a first order polynomial. 
The beam-energy normalized electron momentum, $y$,  
is parameterized by a second order polynomial for the signal, $\tau$ pairs, and 
$\mu$ pairs when one muon decays in flight. For radiative $\mu$ pairs the shape of $y$ 
is approximated by Gaussian with a long asymmetric tail whose mean depends on $E/p$. 
We approximate $dE/dx$ shapes by Gaussians.  
The signal $x$ shape is approximated by a Gaussian 
with the resolution $\sigma(x) = 0.86\% \pm 0.03\%$, 
which we measured using radiative $\mu$ pairs. 
We studied the performance of our fitting method by mixing signal toy MC events with 
calibration data. No biases were observed in these studies. 
We also verified our results by rejecting events 
where the signal electron and muon candidates are back to back. 
Such selection efficiently suppresses the $\mu$-pair backgrounds 
but lowers the sensitivity to the searched-for LFV signal. 
To further verify the analysis presented in this Letter 
we performed a one-dimensional ML fit of the $x$ distribution 
for events remaining after this selection, 
obtaining results consistent with the main analysis 
but with lower efficiency and reduced significance. 

Systematic uncertainties in our analysis arise from several sources. 
The largest contributions to the error on the efficiency come from the 
trigger (5\%), 
event selection (4\%), 
track reconstruction (3\% for two tracks), 
muon identification (2\%), 
online event preselection (2\%), 
signal MC statistics (2\%), 
software trigger (1\%), 
and 
electron identification (1\%) uncertainties. 
The overall systematic error on the efficiency is 8\%. To verify this error estimate we measured 
the partial cross section for $\tau$-pair production in the region $0.65 \le x \le 0.95$ 
using calibration data where no signal and no contamination from $\mu$ pairs are expected. 
Properly scaled up to the total cross section for $\tau$-pair production at 5~GeV, 
our measurement agrees with the expected 0.92~nb within 4\%, 
while the statistical uncertainty of this measurement is 5\%. 

The uncertainty in the $y$ shape and in the efficiency of $y$ region selection for the signal 
are determined by the uncertainty in $\tau$ polarization. 
The polarization of $\tau$ is well-defined for QED processes 
but is model-dependent for BSM contributions. 
The efficiency of the $V+A$ ($V-A$) hypothesis, 
when the electron from $\tau$ decay is boosted forward (backward), 
is 3\% lower (higher) than in the case of an unpolarized $\tau$. 
We use the unpolarized $\tau$ efficiency in the analysis and 
estimate the systematic error in the efficiency of the $y$ region selection 
that arises from $\tau$ polarization uncertainty to be 3\%. 
Significantly larger systematic errors 
(up to 15\%) are associated with the uncertainties in PDF shape parameters 
in ML fitting. To convert signal yields to LFV branching fractions 
we also take into account the 2\% uncertainty in $\Upsilon$ statistics. 

To determine parametric dependence of the likelihood function 
on the signal yield (and LFV branching fraction), 
we integrate the likelihood function over the other three fit parameters, 
{\it i.e.}, the numbers of background events. 
We take the uncertainties in PDF shape parameters 
into account by performing 1000 ML fits for each data sample using 
the PDF shape parameters determined from $\Upsilon(4S)$ and continuum data 
but varied according to Gaussian uncertainties in their values in each fit. 
In addition, to obtain the likelihood distribution for the LFV branching fraction 
we vary the efficiency and the number of $\Upsilon$ mesons in each of these fits 
according to their Gaussian uncertainties. 
The resulting distribution of the likelihood function is the sum of such individual distributions of likelihoods, 
each obtained with its own set of PDF shape parameters, the efficiency and the number of $\Upsilon$ decays. 
This technique takes into account the systematic error on the LFV branching 
fraction arising from the uncertainties in the PDF shape parameters 
and results in widening the likelihood distribution. 

\begin{figure}[h]
\includegraphics*[width=6.50in]{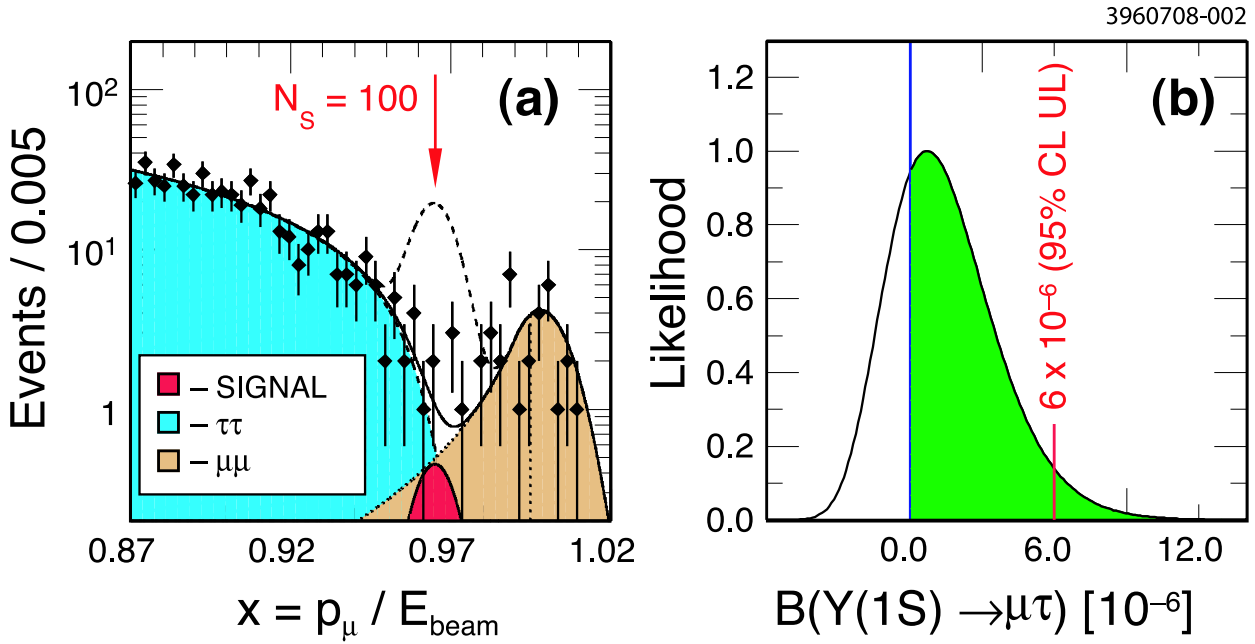}
\caption{
(a) The binned $x$ projection of the results of the ML fit to $\Upsilon(1S)$ data (points with the error bars). 
Solid line indicates the result of the fit, shaded areas show $\tau$-pair, $\mu$-pair and signal contributions to the fit. 
Dashed line shows the hypothetical signal of 100 LFV events superimposed on the result of the fit. 
(b) The distribution of the likelihood function versus branching fraction for LFV decay $\Upsilon(1S) \to \mu \tau$. 
} 
\label{fig_prl_2}
\end{figure}

Our largest signal sample with relatively smaller QED background is $\Upsilon(1S)$ data. 
We show the binned $x$ projection of the results 
of our four-dimensional unbinned ML fit to this sample 
in Fig.~\ref{fig_prl_2}(a). 
The final distribution of the likelihood versus LFV branching fraction 
for leptonic decay $\Upsilon(1S) \to \mu \tau$ is shown in Fig.~\ref{fig_prl_2}(b). 
To estimate the 95\% CL Bayesian UL on this branching fraction 
we integrate the likelihood function for positive ({\it i.e.} physical) values of the branching 
fraction and find the value that correspond to 95\% of the area. 
We apply the same technique to the $\Upsilon(2S)$ and $\Upsilon(3S)$ data and show our results for the 
95\% CL ULs on the branching fractions for LFV decays of $\Upsilon$ mesons in Table~\ref{table_prl}. 

\begin{table}
\begin{center}
\begin{tabular}{cccc}
\hline
\hline
                 & $\Upsilon(1S)$ & $\Upsilon(2S)$ & $\Upsilon(3S)$ \\
\hline
Mass (${\mathrm{GeV}/c^2}$)                                         & 9.46    & 10.02   & 10.36 \\
$N$ decays (millions) & $20.8$ & $9.3$ & $5.9$ \\ 
${\Gamma}(\Upsilon \to \mu\mu)$  (keV)                              & 1.252   & 0.581   & 0.413 \\ 
${\Gamma}(\Upsilon)$             (keV)                              & 53.0    & 43.0    & 26.3  \\ 
${\cal B}(\mu\mu)$  ($\times 10^{-3}$)                              & 23.6    & 13.5    & 15.7  \\ 
${\cal B}(\mu\tau)$ (95\% CL UL, $\times 10^{-6}$)                  & 6.0     & 14.4    & 20.3 \\
${\cal B}(\mu\tau)/{\cal B}(\mu\mu)$ (95\% CL UL, $\times 10^{-3}$) & 0.25    & 1.1     & 1.3 \\ 
$\Lambda$ (95\% CL LL, TeV, $\alpha_N=1.0$)    & 1.30     & 0.98    & 0.98   \\ 
\hline
\hline
\end{tabular}
\smallskip
\caption{
Information necessary to interpret our results 
in terms of BSM physics scale $\Lambda$ and coupling $\alpha_N$. 
We assume lepton universality and use our results for 
dielectron partial widths of $\Upsilon$ mesons \cite{cleo_dilepton}. 
Full widths are according to the PDG summary \cite{PDG}. 
} 
\label{table_prl} 
\end{center}
\end{table}

Effective field theory allows one to relate the dilepton and LFV branching fractions 
of $\Upsilon$ mesons to the scale $\Lambda$ of LFV BSM physics \cite{silagadze,sher} using 

\begin{equation}
\frac{\Gamma(\Upsilon(nS) \to \mu \tau)}{\Gamma(\Upsilon(nS) \to \mu \mu)} = \frac{1}{2 e_b^2} \left( \frac{\alpha_N}{\alpha} \right)^2 \left( \frac{M({\Upsilon(nS)})}{\Lambda} \right)^4, 
\label{eq_2}
\end{equation}

\noindent where $e_b$ is the charge of the $b$ quark, 
$M({\Upsilon(nS)})$ is the mass of vector meson $\Upsilon(nS)$
and $\alpha$ is the fine structure constant. 
We show 95\% CL lower limits (LL) on the BSM energy scale $\Lambda$ 
assuming $\alpha_N=1$ in Table~\ref{table_prl}. This table also shows other quantities 
necessary for estimating $\Lambda$. 

To estimate the lower limit on the scale of BSM physics 
and to produce the exclusion plot of $\Lambda$ versus $\alpha_N$ 
we combine our signal datasets by taking the product of 
individual likelihood functions obtained for each dataset 
before taking into account the systematic errors associated with 
the uncertainties in the overall reconstruction and trigger efficiency, 
PDF shape parameters 
and $\Upsilon$ statistics. 
In the product of the likelihood distributions 
each distribution is represented by 

\begin{equation}
\frac{\alpha^2_N}{\Lambda^4} = \frac{{\cal B}(\Upsilon(nS) \to \mu \tau)}{{\cal B}(\Upsilon(nS) \to \mu \mu)} 
\frac{2 e_b^2 \alpha^2}{(M({\Upsilon(nS)}))^4} 
\label{eq_3}
\end{equation}

\begin{figure}[h]
\includegraphics*[width=6.50in]{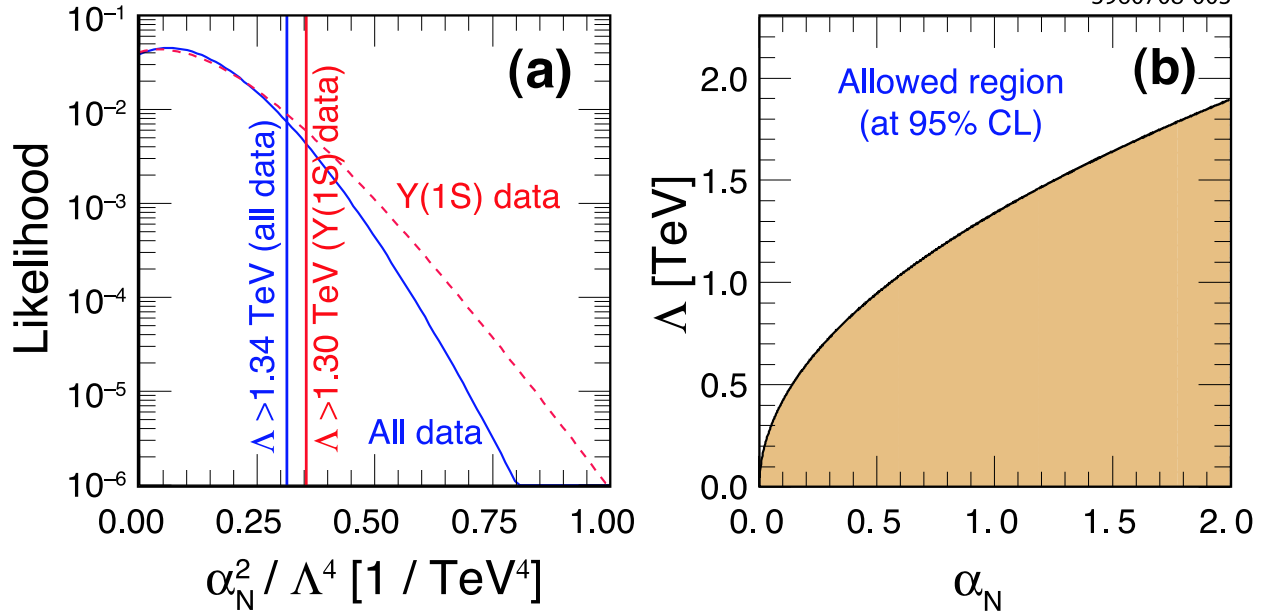}
\caption{(a) The distributions of the likelihood functions versus 
$\alpha_N^2 / \Lambda^4$ (95\% CL ULs are shown assuming $\alpha_N=1$) and 
(b) the exclusion plot for $\Lambda$ versus $\alpha_N$.} 
\label{fig_prl_3}
\end{figure}

\noindent We show the resulting combined likelihood function in Fig.~\ref{fig_prl_3}(a). 
We use this figure to estimate the 95\% CL LL on the scale of BSM physics 
and to prepare the exclusion plot shown in Fig.~\ref{fig_prl_3}(b). 
In Fig.~\ref{fig_prl_3}(a) we show the 95\% CL LLs obtained separately 
with $\Upsilon(1S)$ and, also with all three signal data samples combined. 

The improvement from combining all signal data samples is small 
($\Lambda > 1.34$~TeV using all data
as compared to 
$\Lambda > 1.30$~TeV using the $\Upsilon(1S)$ data), 
because all three samples correspond approximately to the same amount of the integrated 
$e^+e^-$ luminosity and contain similar numbers of background QED events. 
The larger cross section for the production of $\Upsilon(1S)$ 
makes this sample dominate our results for $\Lambda$. 
The slightly more (less) restrictive limits on LFV branching fractions (by 3\%) and 
$\Lambda$ (by 1\%) could be obtained assuming pure $V-A$ ($V+A$) BSM interaction 
for which the efficiency is 9.2\% (8.6\%). Our interpretation of the LFV results from 
the BES experiment \cite{bes_lfv}, $\Lambda > 0.49$~TeV at 95\%~CL, should not be compared 
with our results directly, because these two analyses probe different operators. 
Finally, the lower limits on $\Lambda$ estimated \cite{sher} from the decays of $B$ mesons 
are much more constraining, of the order of hundreds of TeVs, 
than the estimate obtained in our analysis. 
However, such analyses probe non-diagonal operators, 
where the source of possible BSM contribution 
is not necessarily the same as in the analysis presented in this Letter. 

To conclude, 
we searched for leptonic decays $\Upsilon(nS) \to \mu \tau$ ($n=1,2$~and~$3$) 
predicted by various LFV BSM scenarios that would break  
the accidental lepton flavor symmetry of the SM. 
We estimate 95\%~CL ULs on ${\cal B}(\Upsilon(nS) \to \mu \tau)$ 
to be 6.0, 14.4 and 20.3 for $n=1,2$~and~$3$, 
respectively, units of $\times 10^{-6}$. 
In the framework of effective field theory 
we probed the contribution from the operators 
$(\bar{\mu} \Gamma_\mu \tau)(\bar{b} \gamma^\mu b)$ and 
interpret our results in terms of the exclusion plot 
for the energy scale $\Lambda$ of some new BSM 
interaction and the strength of its effective LFV coupling 
and, assuming $\alpha_N=1$, estimate 
the 95\% CL LL on $\Lambda$ to be 1.34~TeV. 

We gratefully acknowledge the effort of the CESR staff 
in providing us with excellent luminosity and running conditions. 
We would like to thank Georges Azuelos, 
Ilya Ginzburg, Adam Leibovich, Marc Sher 
and Arkady Vainshtein for discussions of 
the symmetries in physics 
and various LFV BSM scenarios. 
This work was supported by 
the A.P.~Sloan Foundation,
the National Science Foundation,
the U.S. Department of Energy, and
the Natural Sciences and Engineering Research Council of Canada.

\end{document}